\documentclass[a4paper,12pt]{article}
\usepackage[english]{babel}
\usepackage{amsmath,graphicx}
\begin{document}

\title{Variable-lattice model of multi-component systems. 1. General consideration}
\author{A.Yu. Zakharov$^{\dag}$, A.A. Schneider$^{\dag}$ \\
$^{\dag}$ Novgorod State University\\
Novgorod the Great, 173003, 
Russia\\  
e-mail: anatoly.zakharov@novsu.ru\\  \\
A.L. Udovsky$^{\ddag}$ \\
A.A. Baykov Institute Metallurgy and Material Science\\
Moscow, 119991, Russia }

\date{}
\maketitle 
\begin{abstract}
The paper contains a development of the previously proposed generalized lattice model (GLM). In contrast to usual lattice models, the difference of the specific atomic volumes of the components is taken in account in GLM. In addition to GLM, the dependence of the specific atomic volumes on local atomic environments taken into account in new variable-lattice model (VLM). Thermodynamic functions of multi-component homogeneous phases in the VLM are obtained. Equations of equilibrium between gaseous and condensed phases are derived.
\end{abstract}

\section{Introduction}

The generalized lattice model (GLM) of multicomponent condensed systems (such as solid or liquid solutions) was proposed in paper\cite{ZT} and developed in~\cite{ZZL,ZL,Z1,Z2}. In contradistinction to usual lattice models (see for example~\cite{Prig,Smir,Gray,Isr}), the GLM takes into account the following essential factors: 
\begin{enumerate}
    \item The short-range inter-atomic repulsions. These repulsions are not identical for different pair of atoms, therefore it is impossible to take into account the repulsions by means of an ideal lattice introduction.
    \item The presence of the local fields due to the long-range parts of the inter-atomic potentials. These fields have the essential influence on both equilibrium properties and non-equilibrium processes on the corresponding space scales;
    \item Existence of some non-equilibrium degrees of freedom due to colossal time of their relaxations. Real condensed systems are, as a rule, partially equilibrium systems.
    \item Existence of essential inhomogeneities due to both difference of the atomic sizes of the components and presence of some frozen degrees of freedoms. It is well known, the inhomogeneities scales can be arbitrary up to atomic sizes.
\end{enumerate}

Short-range and long-range parts of the inter-atomic potentials in condensed matter play essentially different roles. Short-range repulsions prevent the collapse of the particles into multi-particle conglomerates and lead to some restriction on the local densities of the components. Long-range parts of the inter-atomic potentials generate the local force fields.

Short-range repulsions in the GLM taken into account by means of the packing condition 
\begin{equation}\label{pack}
    \sum_{i=0}^m \, \omega_i n_i \left( \mathbf{r} \right) - 1 = 0,
\end{equation}
where $\omega_i$ and $ n_i \left( \mathbf{r} \right)$ are the specific atomic volume and local density  of $i$-th component,  respectively, $m$ is number of the components, index $i=0$ corresponds to vacancies as one of the components with equal status in the system. The only difference between vacancies and other (``real'') components is absence direct interaction vacancies with any kind of the particles. 

The Helmholtz free energy functional of the GLM contains has the following form
\begin{equation}
\label{Helm}
\begin{array}{r}
{\displaystyle F \left(T, V, \ldots, \left\{ n_s(\mathbf{r}) \right\}, \ldots  \right)  =\frac{1}{2}\sum_{i,j=1}^{m}\, \iint K_{ij}(\mathbf{r}-\mathbf{r}')\,
n_i(\mathbf{r})\, n_j(\mathbf{r}') \,d\mathbf{r} \,d\mathbf{r}'  }\\
{\displaystyle +T\sum_{i=0}^{m}\int n_i(\mathbf{r}) \ln\left(\frac{n_i(\mathbf{r})}{n(\mathbf{r})}\right)d\mathbf{r},   }    
\end{array}
\end{equation}
where $K_{ij}(\mathbf{r}-\mathbf{r}')$ is the long-range part of the inter-atomic potential of particles $i$-th and $j$-th components, $n_i\left( \mathbf{r} \right)$ is the local density of $i$-th component, $T$ is the absolute temperature in energy units,
\begin{equation}\label{sumn}
n(\mathbf{r})=\sum_{i=0}^{m}\, n_i(\mathbf{r})
\end{equation}
is the summarized density of all the components including the vacancies with their local density $n_0(\mathbf{r})$.

In addition to the Helmholtz free energy functional~(\ref{Helm}) and packing condition~(\ref{pack}), the GLM contains the setting of the total particles numbers of the components
\begin{equation}\label{Ni}
    \int\, n_i(\mathbf{r})\, d\mathbf{r} - N_i = 0.
\end{equation}

Thus, the GLM defined in application to a concrete system by the proper volumes of the components $\omega_i$ and by their long-range parts $K_{ij}(\mathbf{r})$ of the inter-atomic potentials. The equations for the equilibrium distributions can be find from the conditional minimum of the Helmholtz free energy functional~(\ref{Helm}) at the conditions~(\ref{pack}) and~(\ref{Ni}). This way was realized in papers~\cite{ZT,ZZL,ZL,Z2} for both equilibrium states and transport processes.

But, it is very doubtful that the values of proper volumes of the components do not depend on their environment. For example, the vacancies in cooper, in iron, and in lead have not the same volumes. Much more complex the question on vacancies volume is for multicomponent system. 

The aim of present paper is further development and improvement of the GLM with account of the component specific volumes dependence on their local environment.

\section{Basic definitions of the variable-lattice model}

\subsection{General relations}

The most essential imperfection of the GLM due to independence of the components specific atomic volumes $\omega_i$ ($i=0,\, 1,\, \ldots,\, m$) on their environment can be corrected in following way. Let us assume that the proper volume for every of the components is some functional of the distributions of the components
\begin{equation}\label{omega}
\omega_i(\mathbf{r})=\omega_i(\{n_1(\mathbf{r}),...n_m(\mathbf{r})\}).
\end{equation}
Then the packing condition~(\ref{pack}) has the following form
\begin{equation}\label{pack2}
 \sum_{i=0}^m \, \omega_i(\{n_1(\mathbf{r}),...n_m(\mathbf{r})\})\, n_i \left( \mathbf{r} \right) - 1 = 0.    
\end{equation}

Thus, the variable-lattice model is defined by the following relations:
\begin{enumerate}
    \item The Helmholtz free energy functional $F$~(\ref{Helm}), containing the long-range parts $K_{ij}\left( \mathbf{r} \right)$ of the inter-particle potentials;
    \item The packing condition~(\ref{pack2}), containing the short-range inter-particle repulsions with account of the environment influence; this condition holds at every point $\mathbf{r}$ of the system;
    \item The setting of the total particles number for every of the components~(\ref{Ni}).
\end{enumerate}

\subsection{Specific atomic volumes of the components}
In general case, the influence of the environment on the specific atomic volume of the components can be describes by the influence functional. The same particles in the different space points can have different specific atomic volumes because of their environments difference. Any infinitesimal change $\delta n_i\left( \mathbf{r} \right)$ of components distribution in vicinity of the point $\mathbf{r}$ leads to change of the specific atomic volumes in this point 
\begin{equation}\label{ome-i}
    \delta \omega_i \left( \mathbf{r} \right) = \sum_{j=0}^m\, \int \omega_{ij}\left(\mathbf{r} - \mathbf{r}' \right) \delta n_j \left( \mathbf{r}' \right)\, d\mathbf{r}', \quad (i=0,1,\ldots, m)
\end{equation}
where 
\begin{equation}
\label{ome-ij}
\omega_{ij}\left(\mathbf{r} - \mathbf{r}' \right) = \frac{\delta \omega_i \left( \mathbf{r} \right)}{\delta n_j \left( \mathbf{r}' \right)}
\end{equation}
is the variational derivative of the functional $\omega_i \left( \mathbf{r} \right)$ with respect to $n_j \left(\mathbf{r}' \right)$. 
Assume the functions $\omega_{ij}\left(\mathbf{r} - \mathbf{r}' \right)$ are very localized, i.e.
\begin{equation}
\label{omega ij}
\omega_{ij}\left(\mathbf{r} - \mathbf{r}' \right) = \omega_{ij} \, \delta \left(\mathbf{r} - \mathbf{r}' \right),
\end{equation}
where 
$\omega_{ij}$ are the constants (influence coefficients), $\delta\left(\mathbf{r} \right)$ is the Dirac delta-function.
Thus, suppose the connection between local specific atomic volumes and distribution of the components in the following form
\begin{equation}
\label{ome-full}
\omega_{i}(\mathbf{r}) = \omega_{i}^{(0)} + \sum_{j=0}^{m}\omega_{i j}n_j(\mathbf{r}),
\end{equation}
where $\omega_{i}^{(0)}$ are some ``bare'' specific volumes. All the values $\omega_{i}^{(0)}$, $\omega_{ij}$ are the parameters of the variable-lattice model.

\section{Homogeneous multi-components phases}
For a homogeneous $n_i\left(\mathbf{r} \right) = \textrm{const}_i = n_i$ phase the specific atomic volumes of the constituents have the following connections with this phase composition
\begin{equation}\label{omegai}
\omega_{i}=\omega_{i}^{(0)} + \sum_{j=0}^{m}\, \omega_{i j}\, n_j.
\end{equation}
The the packing condition transforms to the following relation:
\begin{equation}\label{pack1}
\sum_{i=0}^{m}\, n_i\left(\omega_{i}^{(0)} + \sum_{j=0}^{m}\omega_{i j}n_j\right) - 1 = 0.
\end{equation}

The configuration term in Helmholtz free energy functional~(\ref{Helm}) for homogeneous system expresses via integral of the long-range parts of the inter-atomic potentials 
\begin{equation}\label{nikij}
\begin{array}{r}
{\displaystyle    \iint K_{ij}(\mathbf{r}-\mathbf{r}')\,
n_i(\mathbf{r})\, n_j(\mathbf{r}') \,d\mathbf{r} \,d\mathbf{r}' = V\, n_i\, n_j\, K_{ij}^{(0)} },
\end{array}
\end{equation}
where
\begin{equation}\label{Kij0}
K_{ij}^{(0)} = \int K_{ij}(\mathbf{r})\, d\mathbf{r}
\end{equation}
is some integral characteristics of the long-range parts of the inter-atomic potential $ K_{ij}(\mathbf{r})$.

Thus, the Helmholtz free energy~(\ref{Helm}) has the following form:
\begin{equation}\label{Helm2}
F =\frac{1}{2}\,\sum_{i,j=1}^{m}K_{ij}^{(0)}\,\frac{N_i N_j}{V} + T\,\sum_{i=0}^{m}\,N_i\, \ln \left( \frac{N_i}{N}\right).
\end{equation}
Let us find the pressure $P$ and chemical potentials $\mu_i$ of the components in homogeneous phases using this expression.

\subsection{Pressure}

Any change of the volume $V$ at fixed values of the particles numbers $N_1$, $N_2$, \ldots, $N_m$  leads to change of the vacancies numbers $N_0$,\footnote{or to their specific volumes changes; note, the volume change generates change of the densities, and the last leads to the specific volumes changes; all of these connections are taken into account.} hence the pressure $P$ is defined as:
\begin{equation}\label{press}
P = -\left(\frac{\partial F}{\partial V}\right)_{T,N_1,\cdots,N_m} = -\left(\frac{\partial F}{\partial V}\right)_{T,N_0,N_1,...,N_m}
-\left(\frac{\partial F}{\partial N_0}\right)_{V,T,N_1,...,N_m}\left(\frac{\partial N_0}{\partial V}\right)_{T,N_1,...,N_m}.
\end{equation}
As a result we have
\begin{equation}\label{press1}
P=\frac{1}{2}\sum_{i,j=1}^{m}K_{ij}^{(0)}n_in_j+T\,
B\left(n_i, \omega_i^{(0)}, \omega_{ij} \right)\,
\ln {\left(\frac{n}{n_0}\right)}.
\end{equation}
where 
\begin{equation}\label{B}
B\left(n_i, \omega_i^{(0)}, \omega_{ij} \right) = \frac{1+\sum\limits_{i,j=0}^{m}\, \omega_{ij}\,n_i\, n_j}{\omega_0^{(0)} +\sum\limits_{j=0}^{m}\, (\omega_{0j}+\omega_{j0})\, n_j}
\end{equation}

\subsection{Chemical potentials}

Any change of the components numbers in the system \textbf{at fixed volume} leads to compensating change of the vacancies number. Hence, the chemical potentials of the components defined by the following way
\begin{equation}\label{chem-pot}
\begin{array}{r}
{\displaystyle  \mu_i = \left(\frac{\partial F}{\partial N_i}\right)_{T,V,\{N_s\}_{s\neq i}}
=\left(\frac{\partial F}{\partial N_i}\right)_{T,V,N_0,\{N_s\}_{s\neq i}}  }\\  \\
{\displaystyle  +\left(\frac{\partial F}{\partial N_0}\right)_{T,V,N_i}
\left(\frac{\partial N_0}{\partial N_i}\right)_{T,V\{N_s\}_{s\neq i}}. }    
\end{array}
\end{equation}
Direct calculations with account of the relations
\begin{equation}\label{dn0-dni}
\left(\frac{\partial N_0}{\partial N_i}\right)_{V,\{N_s\}_{s\neq i}} =
\frac{\omega_i^{(0)} + \frac{1}{V}\, \sum\limits_{j=0}^{m}(\omega_{ij}+\omega_{ji})N_j}{\omega_0^{(0)} + \frac{1}{V}\sum\limits_{j=0}^{m}(\omega_{0j}+\omega_{j0})N_j}
\end{equation}
(it follows from the packing condition~(\ref{pack1})) lead to the following expression for the components chemical potentials:
\begin{equation}\label{chem-pot1}
\mu_i=\sum_{j=1}^mK_{ij}^{(0)}n_i + T\left[\ln \left(\frac{n_i}{n}\right)-
A_i\left(n_i, \omega_i^{(0)}, \omega_{ij} \right)\, \ln\left(\frac{n_0}{n}\right)\right],
\end{equation}
where 
\begin{equation}\label{A-i}
A_i\left(n_i, \omega_i^{(0)}, \omega_{ij} \right) = \frac{\omega_i^{(0)} + \sum\limits_{j=0}^{m} (\omega_{ij}+\omega_{ji})n_j}{\omega_0^{(0)} + \sum\limits_{j=0}^{m}(\omega_{0j}+\omega_{j0})n_j}.
\end{equation}

\subsection{Thermodynamic functions of condensed and gaseous phases}
Let us consider the limiting cases of both gaseous and condensed homogeneous phases. These cases result from the general relations for pressure~(\ref{press1}) and chemical potentials~(\ref{chem-pot1}). To express the pressure $P$ and chemical potentials $\mu_i$ as functions of the system composition and other natural variables, we should first express the densities $n_i$ via atomic concentrations $x_i$, and then exclude the vacancies.

Let us introduce the atomic concentrations of the components by the definition\begin{equation}\label{x-i}
x_i=\frac{n_i}{     \sum\limits_{j=1}^m\,n_j}.
\end{equation}
Hence we obtain
\begin{equation}\label{ni-xi}
n_i=\frac{1 - \omega_0\,n_0}{(\omega,x)}\ x_i,
\end{equation}
where 
\begin{equation}
\label{omega-x}
(\omega,x) = \sum_{i=1}^m\, \omega_i\, x_i\, .
\end{equation}

Substituting (\ref{ni-xi}) into (\ref{press1}), we obtain the expression for pressure via atomic concentrations
\begin{equation}\label{P-x}
P=\frac{ \left(1-\omega_0 n_0 \right)^2}{(\omega,x)^2} \sum_{i,j=1}^m\, K_{ij}^{(0)}\, x_i \,x_j + B\, T\, \ln\left( 1 + \frac{1-\omega_0 n_0}{n_0\,(\omega,x)}\right),
\end{equation}
where $B$ defined by~(\ref{B}).

By the similar way, we obtain the expressions for the chemical potentials
\begin{equation}\label{mu-x}
\begin{array}{r}
{ \displaystyle  \mu_i = \frac{1-\omega_0 n_0}{(\omega,x)}\ \sum_{j=1}^m\, K_{ij}^{(0)}\,x_j +
T\biggl[\ln x_i+\ln \left(\frac{1-\omega_0 n_0}{1-\omega_0 n_0+n_0(\omega,x)}\right)  }\\
{ \displaystyle  
- A_i\ln\left(\frac{1- \omega_0 n_0+n_0(\omega,x)}{n_0(\omega,x)}\right)\biggr],
}    
\end{array}    
\end{equation}
where $A_i$ is defined by~(\ref{A-i}). Note, that $A_i$, $B$ in these relations depend on the phase composition.

\subsubsection{Gaseous phase}
Gaseous phases characterized as phases with the small parameter:
\begin{equation}
\label{eps-1}
\epsilon_1 = 1-\omega_0 n_0 \ll 1. 
\end{equation}
The leading terms with respect to $\epsilon_1$ in (\ref{press1}) and (\ref{chem-pot1}) have the following form:
\begin{equation}\label{P-x2}
P^g = \epsilon_1\, T\,B^g\,\frac{\omega_0^g}{(\omega^g,x)},
\end{equation}
and
\begin{equation}\label{mu-x2}
\begin{array}{r}
{\displaystyle  \mu_i^g = \frac{\epsilon_1}{(\omega^g,x)} \sum_{j=1}^mK_{ij}^{(0)}\,x_j  +
\frac{\epsilon_1\, T}{(\omega^g,x)}\left[\omega_0^g A_i^g-\omega_0^g+(\omega^g,x)\right]  }\\
{\displaystyle + T\ln \left(\frac{\epsilon_1\, x_i\, \omega_0^g}{(\omega^g,\, x)}\right),  }    
\end{array}
\end{equation}
where index ``$g$'' indicates that these relations hold for the gaseous phase.

Excluding the parameter $\epsilon_1$ in relations~(\ref{P-x2}) and~(\ref{mu-x2}), we obtain the chemical potentials as a function of the composition, external pressure and temperature for the gaseous phase:
\begin{equation}\label{mu-PTx}
\begin{array}{r}
{\displaystyle \mu_i^g \left(P, T, x  \right) =  T\ln\left(\frac{P^gx_i}{TB^g}\right)   }\\ 
{\displaystyle + \frac{P^g}{B^g\omega_0^g} \left\{  \frac{1}{T}{\sum\limits_{j=1}^mK_{ij}^{(0)}x_j} + 
\left[\omega_0^gA_i^g-\omega_0^g + (\omega^g,x)\right] \right\}.  }    
\end{array}
\end{equation}

\subsubsection{Condensed phase}
Condensed phases characterized as phases with the small parameter:
\begin{equation}\label{eps-2}
\epsilon_2 = n_0\,\omega_0\, \ll\, 1.
\end{equation}
This parameter is a volume fraction of the vacancies in the system.

The leading terms with respect to $\epsilon_2$ in~(\ref{P-x}) and~(\ref{mu-x}) are
\begin{equation}\label{Press-x}
P^c=\frac{1}{2 (\omega^c,x)^2} \sum_{i,j=1}^{m}\,K_{ij}^{(0)}\, x_i\, x_j-T\,B^c\, \ln \left( \frac{\epsilon_2\,(\omega^c,x)}{\omega_0^c} \right)
\end{equation}  
and
\begin{equation}\label{chem-pot-x}
\begin{array}{r}
{\displaystyle  \mu_i^c = -T\,A^c\,\ln(\epsilon_2)  }\\ \\
{\displaystyle   + \left[T\,\ln x_i - T\,A_i^c\, \ln\left(\frac{(\omega^c,x)}{\omega_0^c}\right) + 
\frac{1}{(\omega^c,x)}\,\sum_{j=1}^m\, K_{ij}^{(0)}x_j\right]
  }.    
\end{array}
\end{equation}

Excluding the parameter $\epsilon_2$ from these relations, we obtain the expressions for the components chemical potentials in condensed phases via composition, temperature, and pressure:
\begin{equation}\label{mu-PTx2}
\begin{array}{r}
{\displaystyle  \mu_i^c = T\ln x_i + \frac{A_i^c}{B^c}P^c  }\\
{\displaystyle + \frac{1}{(\omega^c,x)}\sum_{j=1}^mK_{ij}^{(0)}x_j    - \frac{A_i^c}{2\,B^c\, (\omega,x)^2}\sum_{i,j=1}^{m} \, K_{ij}^{(0)}\,x_i \,x_j\,.   }    
\end{array}
\end{equation}

\section{Phase equilibrium between condensed and gaseous phases}
Equilibrium between gaseous and condensed phases at given pressure~$P$ and temperature~$T$ provided by equalities of the components chemical potentials in these phases
\begin{equation}\label{equi-1}
\mu_i^c = \mu_i^g, \quad (i=1,\ldots, m)
\end{equation}
Substituting~(\ref{mu-PTx}) and~(\ref{mu-PTx2}) into~(\ref{equi-1}), we obtain the complete system of equations for compositions of the co-existing phases
\begin{equation}\label{equi-2}
\begin{array}{r}
{\displaystyle T\ln\left(\frac{Py_i}{TB^g}\right)
+\frac{P}{B^g\omega_0^g}\left\{\frac{1}{T}\sum_{j=1}^m \, K_{ij}^{(0)}y_j+
\left[\omega_0^gA_i^g-\omega_0^g+(\omega^g,y)\right]\right\}  }\\
{\displaystyle  = T\ln x_i + \frac{A_i^c}{B^c}P^c  + \frac{1}{(\omega^c,x)}\sum_{j=1}^mK_{ij}^{(0)}x_j    - \frac{A_i^c}{2\,B^c\, (\omega,x)^2}\sum_{i,j=1}^{m} \, K_{ij}^{(0)}\,x_i \,x_j\,.   }    
\end{array}
\end{equation}
where $y_i$ are the components concentrations in the gaseous phase, $x_i$ are the concentrations in the condensed phase. This system of equations allows to find the composition of one of the co-existing phases by given composition of other phase at given $P$ and $T$.

\section{Conclusion}
The variable-lattice model of the multi-component systems allows take into account the following peculiarities of the condensed matter:
\begin{enumerate}
    \item The short-range interatomic repulsion, including both essential differences of the specific atomic volumes of the components and dependence of atomic volumes on their environments.
    \item The local force fields, due to long-range parts of the interatomic potentials. 
\end{enumerate}
Thermodynamic functions of homogeneous both gaseous and condensed phases are derived. The equations of phase equilibrium between gaseous and condensed phases for variable-lattice model are obtained.

\end{document}